# Constructed wetlands operated as bioelectrochemical systems for the removal of organic micropollutants

Marco Hartl[a,b], María Jesús García-Galán[a], Victor Matamoros[c], Marta Fernández-Gatell[a], Diederik P.L. Rousseau[b], Gijs Du Laing[b], Marianna Garfí[a], Jaume Puigagut[a,*]

[a] GEMMA - Environmental Engineering and Microbiology Research Group, Department of Civil and Environmental Engineering, Universitat Politècnica de Catalunya·BarcelonaTech, c/ Jordi Girona 1-3, Building D1, E-08034 Barcelona, Spain.

[b] Department of Green Chemistry and Technology, Faculty of Bioscience Engineering, Ghent University. Coupure Links 653, 9000 Gent, Belgium.

[c] Department of Environmental Chemistry, IDAEA-CSIC, c/ Jordi Girona, 18-26, E-08034, Barcelona, Spain.

*\* Corresponding author:*

Tel: +34 93 401 08 98

Fax: +34 93 401 73 57

Email: Jaume.Puigagut@upc.edu

**Abstract**

The removal of organic micropollutants (OMPs) has been investigated in constructed wetlands (CWs) operated as bioelectrochemical systems (BES). The operation of CWs as BES (CW-BES), either in the form of microbial fuel cells (MFC) or microbial electrolysis cells (MEC), has only been investigated in recent years. The presented experiment used CW meso-scale systems applying a realistic horizontal flow regime and continuous feeding of real urban wastewater spiked with four OMPs (pharmaceuticals), namely carbamazepine (CBZ), diclofenac (DCF), ibuprofen (IBU) and naproxen (NPX). The study evaluated the removal efficiency of conventional CW systems (CW-control) as well as CW systems operated as closed-circuit MFCs (CW-MFCs) and MECs (CW-MECs). Although a few positive trends were identified for the CW-BES compared to the CW-control (higher average CBZ, DCF and NPX removal by 10-17% in CW-MEC and 5% in CW-MFC), these proved to be not statistically significantly different. Mesoscale experiments with real wastewater could thus not confirm earlier positive effects of CW-BES found under strictly controlled laboratory conditions with synthetic wastewaters.

1. **INTRODUCTION**

Micropollutants can be defined as organic and inorganic substances of anthropogenic origin, with the potential of causing negative effects for the environment already at very low concentrations, i.e. in the order of micro, nano or pico-grams (Chapman, 1992; Stamm et al., 2016). Organic micropollutants (OMPs) include a large array of substances, such as pharmaceuticals, personal care products (PPCPs), hormones, polycyclic aromatic hydrocarbons (PAHs), polychlorinated biphenyls (PCBs), endocrine-disrupting chemicals (EDCs) or pesticides (Bolong et al., 2009).

The four OMPs investigated in this study, i.e. diclofenac (DCF), ibuprofen (IBU), naproxen (NPX) and carbamazepine (CBZ), were all pharmaceuticals and chosen due to their high occurrence in raw wastewaters and/or their low or moderate removal rates in conventional activated sludge (CAS) wastewater treatment plants (WWTPs) (Gros et al., 2012, 2010; Mamo et al., 2018). So far there are no legal discharge limits for OMPs, but some regulations such as the European Decision 2015/495/EU of 20 March 2015 (amending earlier regulations), which included DCF as well as several antibiotics, hormones and pesticides (Barbosa et al., 2016).

A potentially more efficient and less energy intensive alternative to CAS WWTPs are Constructed Wetlands (CWs), which showed promising results regarding OMPs removal (Ávila et al., 2014b, 2014a; Matamoros et al., 2015). The removal efficiency of OMPs in CWs varies with design, operation and type of CW (e.g. surface, subsurface vertical/horizontal flow) employed. In the case of subsurface horizontal flow (HF) CWs, the removal of OMPs ranges from poor to very efficient, depending on characteristics such as bed depth, media size, loading frequency or potential clogging (Ávila et al., 2014b; Matamoros and Bayona, 2006). Various CW intensification strategies have been developed over the last decades and were also tested for the treatment of OMPs, with promising results especially for biodegradable OMPs, but further research is still needed (Ávila et al., 2014b; Nivala et al., 2019; Zhang et al., 2014).

A relatively recent development in the field of wastewater treatment is based on coupling CWs with bioelectrochemical systems (BES) such as Microbial Fuel Cells (MFCs) (Villaseñor et al., 2013; Yadav et al., 2012) and Microbial Electrolysis Cells (MECs) (Ju et al., 2014) called CW-MFC and CW-MEC, respectively, from here on. MFCs use electrochemically active bacteria (EAB) (also known as exoelectrogens, electrogens, electricegens, exoelectrogenic or anode respiring bacteria) as catalysts in order to produce current from the oxidation of organic/inorganic compounds (Logan et al., 2006). These EAB are able to transfer electrons in and out of their cell in a process called extracellular electron transfer (EET) and need a redox-gradient between the MFC electrodes to produce a current. Such a gradient occurs naturally in CW systems. MECs are a modification of MFCs, with the main difference that an external power source is supplied to control the potential between the electrodes, i.e. the anode and cathode, and thereby they are able to achieve thermodynamically reactions, which are otherwise unfavourable (Rozendal et al., 2006). Another advantage of MECs is that only an additional voltage of 0.2-0.8 V is required for water electrolysis to occur (usually 1.8-3.5 V are required), due to the current produced through the activity of EAB at the anode. Both technologies, MFCs and MECs, are able to use wastewater as a substrate and remove a variety of contaminants in the process, showing promising results for the removal of OMPs (Cecconet et al., 2017).

Up to date, there are only a few publications dealing with the use of CW-MFC for OMP removal, and to the authors' knowledge, there are none on the use of CW-MEC systems. Generally, CW-MFC systems have been described to increase the microbial activity (determined by means of fluorescein diacetate (FDA) hydrolysis) in CW-MFC (Hartl et al., 2019), and some studies showed that CW-MFC enhance microbial community richness and diversity as compared to an open-circuit control (Song et al., 2018; F. Xu et al., 2018a). Earlier studies of CW-BES or BES systems for OMP removal used artificial wastewater, which is advantageous for the study of fundamental processes, but less realistic than real urban wastewater (Li et al., 2019; Pun et al., 2019; Wang et al., 2015). The present experiment used CW meso-scale systems which, despite being unplanted, were intended to give additional information on OMP removal in larger scale CW-BES systems with a more realistic horizontal flow, continuous feeding of real urban wastewater and realistic spiking concentration levels of OMPs. Additionally, to the best knowledge of the authors this is the first publication on OMP removal in CW-MEC, and consequently also the first one to compare OMP removal efficiency of CW-MFC and CW-MEC side by side. To this end, duplicate systems with conventional CW (CW-control), closed-circuit CW-MFC (CW-MFC) as well as CW-MEC (CW-MEC) configuration have been used.

The hypothesis was that CW-MEC and CW-MFC will improve organic micropollutants removal as compared to the CW-control system.

## 2. MATERIALS AND METHODS

### 2.1 General design

For the purpose of this work, six unplanted meso-scale horizontal subsurface flow (HF) systems were used (a duplicate of systems per treatment: two systems per CW-control, CW-MFC and CW-MEC). The setup of these systems is detailed in a previous study (Hartl et al., 2019). Briefly, the systems consisted of a PVC reservoir of ca. 0.2 $m^2$ (55 cm length x 35 cm width) surface area filled up with 4/8 mm granitic riverine gravel. The systems were not planted to avoid an additional influencing parameter and further increase the experimental complexity. Wetted depth was set to be 25 cm. At the inlet and around the drainage of the outlet 7/14 mm granitic riverine gravel was used.

Both, the CW-MFC and CW-MEC were designed as three independent BES (MFCs or MECs), respectively, along the length of each system (see Figure 1). Each BES electrode consisted of a gravel-anode with four stainless steel mesh rectangles (Figure 1, C) (SS marine grade A316L, mesh width=4.60 mm, Øwire=1.000 mm, S/ISO 9044:1999) in series acting as electron collectors (4 cm away from each other). The anode area considered for current density calculations was the surface area (0.04 m$^2$) of the anode. Each metal mesh covered nearly the whole cross-sectional area (0.08 m$^2$) of the CW. Each cathode consisted of a carbon felt mat (Figure 1, D) (1.27 cm thick, with a surface area of 0.03 m$^2$, 99.0% carbon purity). A layer of glass wool was placed underneath the cathodes in order to avoid any oxygen leaking from the cathode down to the anode as recommended elsewhere (Venkata Mohan et al., 2008).

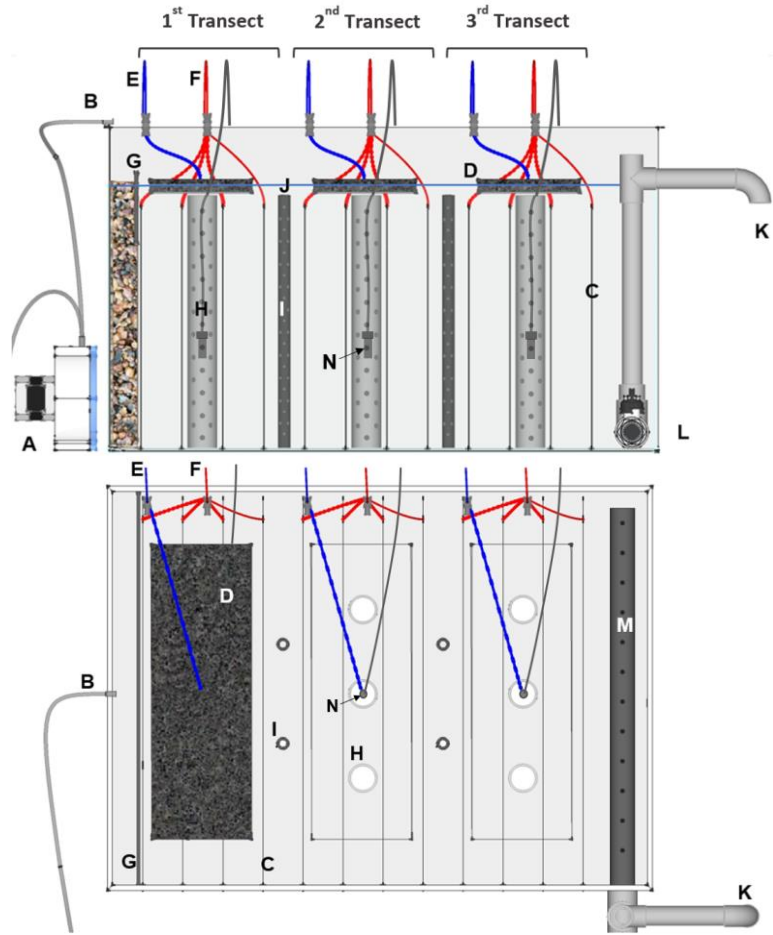

**Figure 1.** Section- (top) and plan-view (bottom) of the CW-BES systems. A: Pump; B: Inflow; C: Anode; D: Cathode; E/F: Anode/Cathode connection to datalogger; G: Inflow barrier to avoid water short-circuiting on surface; H: Gravel core sampling tubes; I: Liquid sampling tubes; J: Water level; K: Standing pipe effluent; L: Drainage; M: Effluent collection tube, N: Reference electrode.

CW-MFC were operated at closed-circuit by connecting the anode and cathode over a wire and a 220 Ω resistor following the recommendations of Corbella and Puigagut (2018). The voltage across the external resistor for each electrode was continuously monitored by means of a datalogger (Campbell Scientific CR1000, AM16/32B Multiplexor). The other two CW-BES systems were operated as CW-MEC by connecting potentiostats to each electrode. CW-MEC systems had the same setup as the CW-MFC but with an additional reference electrode placed near the respective transects anode (Figure 1, N). Each MEC was poised at a potential of 0.3 V vs Ag/AgCl at the anode using a potentiostat (nanoelectra NEV 4). The conventional HF CW-control systems contained no anode metal meshes or electrical connections, but only the cathode carbon felt in order to not confuse physical filtration effects of the carbon felt with BES effects.

## 2.2 Operational conditions

The experimental CWs were mature at the time this work was conducted. The systems of all three treatments were in operation and fed with real urban wastewater already for about 18 months. CW-MFC and CW-control were in the same operation mode throughout the whole time, but CW-MEC systems were operated in CW-MEC mode for the last 9 months leading up to the experiment (the 9 months before that, CW-MEC systems were also run as CW-MFC systems during earlier experiments). During the experiment, the systems were fed with fresh pre-settled urban wastewater every weekday. Influent wastewater was spiked with the target OMPs at a final concentration of 4 µg/L for 4 weeks. Samples for OMP analyses were taken after one week of the start of daily OMP dosing (which represents a bit less than two times the nominal HRT in order to ensure that the OMPs had reached the outlet of the CW during sampling).

Further details on pre-treatment (settling for 3 hours) and operation are given in Hartl et al. (2019). The average hydraulic loading rate (HLR) applied during the experiment was 28 mm/d, resulting in a nominal HRT of 3.6±0.3 days and an average organic loading rate (OLR) of 8.7±2.5 g COD/m² day.

## 2.3 Sampling and analysis

### 2.3.1. Water quality parameters

Eight sampling campaigns for the characterization of conventional wastewater quality parameters were conducted during 12 weeks. These campaigns were conducted already 3 weeks before OMP sampling started, and continued during the OMP sampling period, whereas conventional wastewater samples were taken just before the OMP dosing on weekdays. Conventional wastewater parameters were measured for the influent, after the first and second third of the wetland length, and as also at the effluent. All samples were analysed for total suspended solids (TSS), volatile suspended solids (VSS) and total chemical oxygen demand (COD) according to Standard Methods (APHA-AWWA-WEF, 2012); $NH_4^+$-N, according to Solórzano method (Solórzano, 1969); $NO_2^-$-N, $NO_3^-$-N, $SO_4^{2-}$-S and $PO_4^{3-}$-P by ion chromatography (ICS-1000, Dionex Corporation, USA). Physical parameters such as water temperature, dissolved oxygen (DO) concentration and pH were measured directly in the influent, using portable devices after the first and second transect, as well as in the effluent (EcoScan DO 6, ThermoFisher Scientific, USA and CRISON pH/mV – meter 506, Spain, respectively). Further details are given in Hartl et al. (2019).

### 2.3.2. OMP analysis

High purity standards (>99%) of the parent compounds and the isotopically labelled compounds were purchased from Sigma-Aldrich (St. Louis, MO, USA). Detailed information on their physical and chemical characteristics is given in Table S1 of the Supplementary Information (SI). Standard solutions of the mixtures of the four compounds were made at the appropriate concentrations and used to dope the influent wastewater. Five OMP sampling campaigns were conducted during 3 weeks. Grab samples were taken from the CW influent and effluent sampling points (see Figure 1, points B and K, respectively). All water samples were filtered and processed using an adapted methodology by Matamoros and Bayona (2006). Briefly, 50 mL of influent and 100 mL of effluent samples were filtered (0.7 µm Whatman™ glass microfiber filters GF/F), acidified to pH 2-3 with HCl (0.02M) and spiked with a mixture of surrogate standards to a final concentration of 50 ng L$^{-1}$ (atrazine-d$_5$, mecoprop-d$_3$, tonalide-d$_3$, and dihydrocarbamazepine). Solid phase extraction was then performed, using 200 mg Strata™-X polymeric cartridges from Phenomenex (Torrance, CA, US), previously conditioned with 3 mL of hexane, 3 mL of ethyl acetate, 5 mL of MeOH and 5 mL of acidified milli-Q water. Elution was performed with 10 mL of hexane/ethyl acetate (1:1, v:v). The eluted extract was evaporated under a gentle nitrogen stream to a volume of 100 µL, and triphenylamine was added as an internal standard (20 ng). Finally, vials were reconstituted to 300 µL and analysed by GC-MS/MS as described by Matamoros et al. (2017).

**2.4 Data Analysis**

Contaminant removal efficiencies were calculated on a mass balance basis taking into account the wastewater flow and pollutant concentration. Statistical analyses were conducted for comparison of the relevant parameters COD, $NH_4^+$-N, $SO_4^{2-}$-S, $PO_4^{3-}$-P, pH and current density as well as the OMPs CBZ, DCF, IBU and NPX across the three treatments. Since Shapiro-Wilk tests showed a normal distribution for all data, single-factor analysis of variance (ANOVA) could be used on all the above mentioned parameters. Post-hoc Tukey HSD and Scheffé multiple comparison tests were performed for all parameters but only presented when relevant, i.e. in the case that ANOVA reported significant differences between treatments, thus only in the case of pH. The software used for calculations and statistical analysis was Microsoft® Excel® 2016 and the included Analysis ToolPak add-in.

## 3. RESULTS AND DISCUSSION

### 3.1 Electrical behaviour

Table 1 shows average and maximum measured cell voltages ($E_{cell}$) as well as consequent current and power densities per surface area and anodic compartment volume for CW-MFC treatments in all 3 transects.

**Table 1.** Average, standard deviation and maximum for $E_{cell}$ as well as current density and power density of closed-circuit CW-MFC systems. Note: The surface area of each electrode was used for current density calculations.

| Transect | $E_{cell}$ (mV) | | Current Density per Area (mA/m$^2$) | | Current Density per Volume (mA/m$^3$) | | Power Density per Area (mW/m$^2$) | Power Density per Volume (mW/m$^3$) |
|---|---|---|---|---|---|---|---|---|
| | Avg ± SD | Max | Avg ± SD | Max | Avg ± SD | Max | Avg ± SD | Avg ± SD |
| 1 | 372±119 | 552 | 40±13 | 60 | 183±59 | 273 | 15.0±1.5 | 68±7 |
| 2 | 378±81 | 577 | 41±9 | 62 | 186±41 | 282 | 15.4±0.7 | 70±3 |
| 3 | 372±128 | 711 | 40±14 | 77 | 183±64 | 350 | 15.0±1.8 | 68±8 |

Average current densities per surface area for CW-MFC (all transects considered) resulted in 40 mA/m$^2$. Differences in current density between transects were not statistically significant according to ANOVA. A polarization curve (PC) analysis (see SI, Figure S1) showed that maximum power densities of 30, 11 and 24 mW/m$^2$ in transect 1, 2 and 3 of CW-MFC mode, respectively, were achieved at current densities of 79, 35 and 66 mA/m$^2$, respectively, which is higher than that described by Saz et al. (2018) (ca. 20 mA/m$^2$) under comparable conditions.

The estimated internal resistances derived from the polarization curves were around 108 Ω, 220 Ω and 124 Ω for first, second and third transect, respectively. Principally, the potential maximum power is achieved when internal and external resistances are close to each other (Lefebvre et al., 2011). Coincidentally, the external and internal resistance were exactly the same in transect 2. However, for the current experiment and its primary goal contaminant removal the lower external resistances in transects 1 and 3 could have been beneficial, since lower external resistances increase the generated current and studies have also shown that consequently organic matter removal was increased (Aelterman et al., 2008; Gil et al., 2003; Katuri et al., 2011).

The coulombic efficiency (CE) is the proportion of the produced current to the carbohydrates which are theoretically derived from oxidation, indicated by the change of COD from transect to transect (Scott, 2016). The resulting average CE values amounted to 1.4±2.4%, 9.5±7.6% and -29.4±4.6%, for transects 1, 2 and 3, respectively. Note that CE can have a negative value when COD concentrations were increasing from the influent to the end of transect 1 or from one transect to the other. Generally, it can be assumed that only the CE value measured in transect 1 gives a good indication since not only organic matter from the influent can contribute to the MFC signal but also accumulated organic matter within the gravel bed is a fuel source for MFC (Corbella et al., 2016). Therefore, CE in transect 2 could be higher than transect 1 and transect 3 CE even negative on average. Comparable CW-MFC studies produced CEs from 0.01‰ (Wang et al., 2016b) up to 16.4% (Xie et al., 2018).

Table 2 shows the poised potential and the resulting achieved average current as well as average current and power densities per electrode surface area and anodic compartment volume for each transect in CW-MEC systems.

**Table 2.** Poised potential (at Anode vs. Ag/AgCl reference electrode) as well as resulting average current applied also expressed in current and power density per surface area and volume in CW-MEC (MEC)

| Transect | Poised Potential (V) | Current (mA) | Current Density per Area (mA/m$^2$) | Current Density per Volume (mA/m$^3$) | Power Density per Area (mW/m$^2$) | Power Density per Volume (mW/m$^3$) |
|---|---|---|---|---|---|---|
| | | Avg ± SD | Avg ± SD | Avg ± SD | Avg ± SD | Avg ± SD |
| 1 | 0.3 | 23±11 | 535±263 | 2434±1197 | 161±79 | 730±359 |
| 2 | 0.3 | 10±5 | 223±112 | 1015±510 | 67±34 | 304±153 |
| 3 | 0.3 | 5±3 | 120±74 | 545±334 | 36±22 | 163±100 |

The poised potential of 0.3 V vs. Ag/AgCl reference electrode at the anode, was chosen on the basis of experiences showing that poised potential around this value benefit the growth of electrochemically active bacteria (EAB) genera such as *Geobacter* in mixed bacterial cultures (Fricke et al., 2008; Liu et al., 2008). The average current density in CW-MEC was more than double in transect 1 compared to transect 2, and transect 2 was again roughly double of transect 3, assumingly because the organic matter concentration was decreasing along the flow path through the systems.

The CW-MEC current densities in all three transects were low when compared to other similarly built CW-MEC systems which showed values ranging from 200 to 24500 mA/m$^2$ (Gao et al., 2017; Srivastava et al., 2018; Xu et al., 2017; Zhang et al., 2018). Authors believe that the use of non-conductive media (gravel) as anode material is the main cause of the lower observed current densities in the CW-MEC systems because this increases the internal resistance of the systems.

### 3.2 Removal efficiency of conventional wastewater quality parameters

Results on the removal of conventional contaminants in all three treatments (CW-control, closed-circuit CW-MFC and CW-MEC systems) are summarized in Table 3. All results were obtained during 8 weeks of intensive sampling (5 weeks before the OMP sampling campaigns and the three weeks during the OMP sampling campaign). Data is shown as average mass loading rate at the system inlet (influent), after the first and second transects and effluent, as well as mass removal rate from influent to effluent based on the average mass and percentage. During this period, all systems received continuous flow with an average OLR of 8.7±2.5 g COD/m$^2$.day.

**Table 3. Results for COD, ammonium, sulphate and orthophosphate for CW-control, closed-circuit CW-MFC and CW-MEC systems during the 8 sampling weeks, expressed as average mass loading rate at influent, after first transect, after second transect and effluent as well as removal from influent to effluent based on the average mass removal rate and percentage.**

|  |  | Influent | 1/3 | 2/3 | Effluent | Removal from Influent to Effluent | |
|---|---|---|---|---|---|---|---|
|  |  | (g/m$^2$.d) | | | | (g/m$^2$.d) | (%) |
| COD (n=8) | CW-control | 8.6±2.6 | 4.4±1.9 | 3.7±2.2 | 3.7±1.6 | 4.9±1.4 | 57 |
|  | CW-MFC | 8.9±2.4 | 4.4±2.3 | 3.8±2.3 | 4.0±1.5 | 4.9±0.5 | 55 |
|  | CW-MEC | 8.7±2.5 | 3.9±2.3 | 2.5±1.4 | 2.6±1.0 | 6.1±0.8 | 70 |
| NH$_4$-N (n=7)[a] | CW-control | 1.2±0.4 | 1.0±0.4 | 0.9±0.3 | 1.1±0.3 | 0.1±0.2 | 10 |
|  | CW-MFC | 1.3±0.4 | 1.0±0.3 | 0.8±0.2 | 1.0±0.3 | 0.3±0.2 | 24 |
|  | CW-MEC | 1.2±0.4 | 0.9±0.3 | 0.7±0.2 | 0.9±0.2 | 0.3±0.3 | 28 |
| SO$_4^{2-}$ (n=6)[a] | CW-control | 2.0±1.3 | 0.5±0.5 | 0.4±0.3 | 0.8±0.6 | 1.1±0.9 | 58 |
|  | CW-MFC | 2.1±1.4 | 0.6±0.4 | 0.6±0.4 | 1.1±0.9 | 1.0±0.3 | 48 |
|  | CW-MEC | 2.2±1.4 | 0.8±0.7 | 1.0±0.8 | 1.1±0.9 | 1.1±0.8 | 51 |
| PO$_4^{3-}$-P (n=6)[a] | CW-control | 0.09±0.05 | 0.09±0.06 | 0.09±0.05 | 0.09±0.05 | 0.00±0.03 | 2 |
|  | CW-MFC | 0.09±0.05 | 0.09±0.06 | 0.08±0.05 | 0.08±0.05 | 0.01±0.03 | 7 |
|  | CW-MEC | 0.09±0.05 | 0.07±0.06 | 0.06±0.05 | 0.08±0.05 | 0.01±0.04 | 7 |

[a] Some experimentation weeks could not be considered due to highly diluted influent or technical analysis problems

In contrast to previous studies done on the same experimental systems study (Hartl et al., 2019), ANOVA reported no statistically significant differences for general wastewater quality parameters, with the exception of pH (see SI, Table S2 and S5). NH$_4^+$-N removal was generally low, and decreased towards the end of the study period, which was also observed for COD, although to a lesser extent. It is assumed that ageing and possible partial clogging of the carbon felt cathodes might have limited removal performance of CW-MFC systems.

Indeed, results from MFC studies (Kim et al., 2008; Yan et al., 2012; Yuan et al., 2021) suggested that their single-chamber air-cathode MFC system provided a suitable microenvironment for biological ammonia oxidation whereas the ammonia oxidizing bacteria (AOB) found in the cathode biofilm derived oxygen via diffusion through the cathode. So the observed partial clogging of the cathode could have limited the activity of AOB at the cathode, and as a consequence also have reduced nitrification and $NH_4^+$-N removal. In the case of $NH_4^+$-N, it could also be assumed that due to the system ageing and the accompanying development and establishment of the microbial communities, the systems turned more anaerobic, which in turn would have lowered nitrification rates.

For both COD and $NH_4^+$-N removal, the lack of plants in the meso-scale systems could have had an effect as well on overall treatment efficiencies, since the presence of plants has shown to improve treatment efficiency in HF CWs (Tanner, 2001). $NO_2^-$-N and $NO_3^-$-N were generally below the limit of detection. A recent study showed that planted CW-MFC systems show higher power density and contaminant removal, however, dead plant parts in turn also reduced the power production (Yang et al., 2019).

Generally, $PO_4^{3-}$-P removal efficiency was lower when compared to current literature regarding CW-MFC (Corbella and Puigagut, 2018; Saz et al., 2018; Xu et al., 2018; Yakar et al., 2018) or CW-MEC (Gao et al., 2018; Ju et al., 2014; Zhang et al., 2018) systems. However, many studies were conducted only over a short-time and it is generally known that phosphorus storage in subsurface flow CWs has a finite capacity and therefore removal by sorption normally decreases over time (Kadlec and Wallace, 2009), as could have been the case in this study as the wetlands were operated for about 18 months.

The average values for pH measurements at each sampling point are shown in SI, Table S2. The ANOVA results for influent and average pH values of all sampling points were statistically not significantly different across treatments. However, after the first transect, CW-MEC systems showed a lower pH than other treatments on average, being significantly different from CW-MFC as well as CW-control according to ANOVA, with both, post-hoc Tukey HSD and Scheffé pairwise comparison showing CW-MEC to be either significantly or very significantly different to CW-MFC and CW-control (for more details see SI, Figures S3 and S4). After the second transect, pH values of all three treatments were significantly different from each other according to ANOVA, with CW-MEC showing the lowest pH, followed by a higher pH in CW-MFC and the highest in CW-control (meaning the smallest change since the influent inlet in the system). Post-hoc Tukey HSD and Scheffé pairwise comparison of pH after the second third showed very significant differences between all treatments (for more details see SI, Figures S3 and S4). pH values at the effluent were generally higher than in the previous two transects within the systems, and the difference between treatments was again only statistically different in the CW-MEC systems. Post-hoc Tukey HSD and Scheffé pairwise comparison of pH at the effluent showed CW-MEC to be very significantly different to CW-MFC and CW-control (for more details see SI, Figures S3 and S4).

Changes in pH within the system might affect the activity of bacteria, and influence the charge state as well as hydrophobicity of certain OMPs (Wang et al., 2015). While the measured pH in solution showed some significant differences between treatments, the changes seemed not big enough to alter the charge state and hydrophobicity of the investigated OMPs significantly, especially in the case of CBZ with its high $pK_a$ of 13.9 (see SI, Table S1). However, pH at the micro-scale, e.g. near the cathode or anode, might have changed more drastically, and could have created micro-environments in which charge state and/or hydrophobicity were influenced. Unfortunately, it was not possible to measure these changes in pH on a micro-scale with the presented setup.

## 3.3 Removal efficiency of organic micropollutants

Table 4 shows the removal of the four targeted OMPs for all three treatments (see also SI, Figure S2 for box- and whisker plots).

**Table 4.** Results for OMPs carbamazepine (CBZ), diclofenac (DCF), ibuprofen (IBU) and naproxen (NPX) in CW-control, closed-circuit CW-MFC and CW-MEC systems during the 5 sampling campaigns, expressed as average background, influent and effluent concentration, average mass loading rate at influent and effluent as well as removal from influent to effluent based on the average mass removal rate and percentage. (Concentration variability in the influent concentrations is due to the background concentration of the urban wastewater for each of the compounds).

| OMP (n=5) | Back-ground (µg/L) | Influent (µg/L) | Influent (µg/m².d) | Treatment | Effluent (µg/L) | Effluent (µg/m².d) | Removal (µg/m².d) | Removal (%) |
|---|---|---|---|---|---|---|---|---|
| CBZ | 3.5±2.2 | 5.3±2.2 | 149±61 | CW-control | 4.6±1.4 | 123±41 | 26 | 17 |
|  |  |  |  | CW-MFC | 4.3±1.0 | 116±26 | 33 | 22 |
|  |  |  |  | CW-MEC | 3.7±0.8 | 99±24 | 50 | 34 |
| DCF | 0.6±0.3 | 4.2±1.9 | 137±56 | CW-control | 2.7±1.4 | 73±17 | 65 | 47 |
|  |  |  |  | CW-MFC | 2.2±1.0 | 65±20 | 72 | 52 |
|  |  |  |  | CW-MEC | 2.2±0.8 | 59±16 | 79 | 57 |
| IBU | 12.6±3.6 | 18.6±8.8 | 523±202 | CW-control | 12.0±2.0 | 321±53 | 202 | 39 |
|  |  |  |  | CW-MFC | 12.6±1.7 | 341±40 | 182 | 35 |
|  |  |  |  | CW-MEC | 12.0±2.2 | 320±52 | 202 | 39 |
| NPX | 3.8±0.7 | 10.2±1.4 | 273±29 | CW-control | 7.6±2.4 | 203±62 | 70 | 25 |
|  |  |  |  | CW-MFC | 7.1±2.0 | 191±50 | 82 | 30 |
|  |  |  |  | CW-MEC | 6.1±1.5 | 163±37 | 109 | 40 |

Similar as for the general wastewater parameters, removal differences across treatments were not statistically significant for any of the four compounds (see SI, Table S6). Nevertheless, few tendencies were visible which are discussed further below.

**Carbamazepine**

The CW-control system removal of 17% is in accordance with results of previous studies on treatment capacity in conventional HF CW systems (not operated as BES), reporting removals of 13% (Nivala et al., 2019) and 21% (Matamoros et al., 2017). These results show that CBZ can be removed to a certain degree in HF CWs (supposedly due to anaerobic processes), however, CBZ is not biodegradable in aerobic conditions and therefore VF CWs show lower removal rates (Hai et al., 2011; Jekel et al., 2015; König et al., 2016; Nivala et al., 2019).

The only other study looking at CBZ removal in CWs operated as BES resulted in removal of more than 99% from synthetic wastewater (Pun et al., 2019). However, this system was operated in short-circuit and used a bed of highly porous and electroconductive media (graphitized coke), in which anodic and cathodic processes were uncontrolled (comparable to a CW-MFC but without solid state electrodes or external connection). Their own sorption experiments showed that ca. 30% of the compound was removed solely by abiotic sorption onto the highly porous media. Also in conventional MFC and MEC (poised potential of -0.4 V vs Ag/AgCl at the anode) systems, Werner et al. (2015) identified hydrophobic sorption as the dominant mechanism for CBZ removal, attributing the removal (>80%) mainly to the large anode areas provided by the graphite fibre brushes (material with high sorption propensity) and the attached biofilm. However, graphite has a high sorption propensity as well, unlike the used gravel in the presented study. Although CBZ can actually not be considered hydrophobic (log D of 2.77, see SI, Table S1), it is less polar than the other three tested OMPs, and therefore the contribution of sorption to CBZ removal is potentially higher than in the three other tested OMPs. Generally, CBZ is considered a recalcitrant due to its low removal in conventional CAS, which rarely exceeds 10% (Joss et al., 2005; Zhang et al., 2014).

Despite being non-significant according to the performed ANOVA (see SI, Table S6) – probably the result of working with real wastewater with variable composition – the results obtained for CW-MEC and CW-MFC show a tendency for improvement compared to CW-control (see Table 4 and SI, Figure S2a). There are several processes which could potentially play a role here.

Electrosorption and hydrophobic sorption could have influenced CW-MFC and CW-MEC by offering additional sorption sites at the electrodes and the biofilm, and thereby improved the removal. However, these sorption sites are finite and longer term investigations using BES incorporated in CWs for CBZ removal are suggested. An effect of pH changes (see SI, Table S2) on hydrophobicity and charge in the different treatments is unlikely in the case of CBZ due to the high $pK_a$ of 13.9 (see SI, Table S1).

However, an increase in microbial activity observed in CW-MFC in an earlier study (Hartl et al., 2019), could have led to an improved biodegradation and at least partly explain the tendency for higher removal in CW-MFC and possibly CW-MEC as compared to the CW-control. Although no microbial activity studies in CW-MEC are known to the authors it could be assumed that it is affected in a similar way as in CW-MFC. Further investigation of the microbial communities, especially of CW-MEC, are suggested.

**Diclofenac**

DCF removal of 47% in CW-control was higher than in other publications on conventional HF CW systems, reporting 25% (Nivala et al., 2019) and 19±21% removal (Matamoros et al., 2017). There are no publications yet on DCF removal by CW-MFC or CW-MEC systems. DCF removal rates in the presented CW-MFC were high even when compared to conventional MFC systems fed by synthetic wastewater, which reached only 4-8% in a single-chamber closed-circuit MFC and up to ca. 23% and 45% in the anode and cathode chamber of a double chamber MFC, respectively (Wang et al., 2015). De Gusseme et al. (2012) applied biogenic Pd nanoparticles as a biocatalyst to a conventional MEC (voltage of -0.8 V applied to the circuit) for the catalytic dechlorination of DCF (from synthetic wastewater with 1 mg/L DCF) and achieved full removal while no significant removal was achieved without the use of the nanoparticles. In conventional CW systems (not operated as BES), vertical flow (VF) CW systems are more efficient in removing DCF through aerobic processes, with performances ranging from 50-70% (Ávila et al., 2014a, 2014b; Matamoros et al., 2007; Nivala et al., 2019), while the removal in HF CWs is lower and thought to happen through anaerobic degradation (Ávila et al., 2010). The biological removal of DCF is not fully understood and results are usually very variable (Zhang et al., 2008). DCF is also a recalcitrant (though not as strongly as CBZ), thus removal rates in conventional WWTPs can be also relatively low and variable with elimination values in the range of 7-75% (Zhang et al., 2014).

Although the log $K_{ow}$ of DCF is high with 4.26, it gets deprotonated and becomes highly hydrophilic at the pH range of 6.6 to 7.6 of the presented systems, with a log D of 1.70 to 1.04 (see SI, Table S1), resulting in a low sorption propensity. Given the charge and sorption characteristics of DCF, conventional sorption and pH effects seem unlikely to influence the DCF removal to a great extent. However, in theory, electrosorption at the electrode with opposite charge (i.e. at the positively charged cathode, since DCF has a negative charge, see SI, Table S1) (Kong et al., 2013; Yang et al., 2015) could have resulted in an increased tendency for DCF removal (see Table 4 and SI, Figure S2b). Apart from that, CW-MFCs have been proven to enhance microbial activity (Hartl et al., 2019). Additionally, potential electrolysis of water in CW-MEC could produce oxygen and $H^+$ at the anode and $H_2$ at the cathode. The produced oxygen could increase the aerobic biodegradation of DCF in CW-MEC. However, these effects could not be (statistically) confirmed according to the ANOVA (see SI, Table S6).

**Ibuprofen**

As for all other OMPs, also an ANOVA of IBU removal was not statistically significantly different across treatments (see SI, Table S6), and showed also in relative comparison the smallest differences between treatments with 39% removal in CW-control and CW-MEC, and 35% in CW-MFC systems (see Table 4 and SI, Figure S2c). Anyway, the here reported removal rates were comparable to those found in two exemplary HF CW systems amounting to 28% (Matamoros et al., 2017; Nivala et al., 2019). To the knowledge of the authors, there are no publications yet on IBU removal by CW-MEC systems and just one other publication which currently addresses IBU removal using a CW-MFC; Li et al. (2019) reported IBU removal rates of 82-96% from synthetic wastewater in a CW-MFC, which was 9% higher than their open-circuit control, with 63-79% of the removal happening in the anodic section.

Removal rates in conventional MFC systems reached values of 18-20% in single-chamber closed-circuit systems, and up to ca. 40% and 87% in anode and cathode chambers of a double-chamber MFC, respectively (synthetic wastewater was used) (Wang et al., 2015). In general, IBU is highly hydrophilic and therefore sorption is low, with a log D of 1.16 to 2.10 in the measured pH range (see SI, Table S1). Aerobic conditions favour its biodegradation (Monsalvo et al., 2014; Quintana et al., 2005), hence VF CWs show removal rates above 88% (Ávila et al., 2010; Nivala et al., 2019; Vystavna et al., 2017). This is probably also why plants – known to provide oxygen to the systems via their roots (Kadlec and Wallace, 2009) - improved IBU removal in HF CWs (Li et al., 2016). Removal rates in conventional WWTPs are usually high (41-100%) due to the prevalent aerobic removal mechanisms (Zhang et al., 2014). In general, the authors suggest to confirm the obtained results of all OMPs in planted CWs operated as BES.

In summary, IBU removal was not notably improved through CW-MFC or CW-MEC, although other studies on CW-MFC or conventional MFC were able to achieve that in comparison to control systems. In terms of charge, sorption propensity and biodegradability, IBU has similar characteristics as DCF and NPX, therefore other factors seem to be responsible for the even clearer lack of difference between treatments. Further investigation shall be carried out to confirm and possibly explain the results reported here.

**Naproxen**

The 25% NPX removal in the CW-control was lower than in comparable HF CW systems showing 32% (Nivala et al., 2019) and 66% removal (Matamoros et al., 2017). The short-circuit CW-BES by Pun et al. (2019) removed more than 95% of NPX from synthetic wastewater; only a fraction (13.1-18.5% according to abiotic sorption tests) of that was retained within the material and therefore unrelated to biological activity of bacteria. Removal rates in conventional MFC systems by Wang et al. (2015) reached ca. 12-19% in single-chamber closed-circuit systems and up to ca. 40% and 84% in the anode and cathode of double-chamber MFC, respectively (all using synthetic wastewater).

Sorption of NPX is low, with a log D of 0.61 to -0.18 at the pH range of 6.6-7.6 (see SI, Table S1). Generally, NPX is mainly removed by biodegradation, and preferably under aerobic conditions (Kahl et al., 2017), hence VF CWs show high removal rates above 88% (Ávila et al., 2010; Nivala et al., 2019; Vystavna et al., 2017). Again as for IBU, removal rates in conventional WWTPs are relatively high and in the range of 40-98% (Zhang et al., 2014).

As for DCF, the positive tendencies seen in CW-BES (see Table 4 and SI, Figure S2d) could be possibly due to electrosorption. Also an increase in microbial activity and/or a potential increase in oxygen through electrolysis at the anode could have led to the observed slight NPX removal improvement.

In general, according to Cecconet et al. (2017), BES are theoretically more efficient in removing OMPs which are hydrophobic and positively charged. The former due to the better adsorption onto charged electrodes and the latter due to the better interaction with the negatively charged biofilm. The four OMPs presented in this study are all hydrophilic at neutral pH and either negatively charged (DCF, IBU and NPX), or neutrally charged (CBZ) under the pH range of the systems (see SI, Table S1), which could be additional reasons for the statistically not significant results.

As mentioned above, in the case of DCF and NPX the observed insignificant but yet positive tendencies could be due to electrosorption to the positively charged electrode in in CW-BES. These OMPs are present in the form of charged ions or polar molecules and could therefore have been adsorbed after migrating to the system´s electrode with opposite charge (Kong et al., 2013; Yang et al., 2015). Apart from that, MFCs seem to offer a beneficial environment for the growth of non-electrochemically active bacteria and increasing the metabolic rate of anaerobes due to the artificial presence of an insoluble electron acceptor, i.e. an anode (Fang et al., 2013). Another important factor to consider - apart from charge, sorption effects, microbial activity and direct impact on microbial communities - is the biodegradability of the compound (Wang et al., 2015). The BES itself might have influenced environmental conditions, especially on a micro-scale (e.g. at the electrodes or adjacent pore spaces) changing factors like pH (with statistically significant differences) and DO, which in turn could have indirectly affected microbial communities and their degradation of OMPs in the systems. Unfortunately, as mentioned above it was not possible to measure these parameters on such a small scale in the present study. However, similar studies reported electrolysis at the anode of CW-MEC systems (Gao et al., 2017) which would cause oxygen and hydrogen to be released and consequentially increase aerobic and hydrogen consuming microbial processes which could increase the removal of OMPs which show high removal rates in aerobic processes such as DCF and NPX.

## 4. CONCLUSIONS

The investigation of meso-scale CWs operated as BES (CW-BES) resulted in the following conclusions:

- Contrary to the hypothesis, no statistically significant effect of CW-BES on OMP removal could be found. A potential reason could have been the use of real urban wastewater which is more variable than synthetic wastewater. In general, the authors suggest careful consideration of results based on artificial conditions and recommend continued research with application of real urban wastewater and conditions as realistic as possible.

- However, some tendencies of increased OMP removal were noted in CW-MEC and CW-MFC when compared to CW-control for three out of the four investigated pharmaceuticals, namely CBZ, DCF and NPX with an average increase of 10-17% in CW-MEC and 5% in CW-MFC systems, compared to the CW-control. These tendencies could be due to various reasons such as increased microbial activity, or indirect effects through an electrolysis induced increase of DO and subsequent aerobic degradation (at least in the case of DCF and NPX in CW-MEC mode). Hydrophobic (and electro-) sorption might have played an additional role in the removal of CBZ, and electrosorption effects in the case of DCF and NPX. More long-term observation periods are recommended in order to take into account the inherent limitation of (electro-)sorption sites.

- In contrast to earlier research, no statistically significant removal was found regarding conventional wastewater parameters such as COD and $NH_4^+$-N, potentially due to ageing effects of the systems, especially clogging of cathodes, which could have also influenced the BES performance and consequently OMP removal. Thus, further investigations into the long-term ageing and clogging effects on electrodes are suggested, which would ideally lead to practical recommendations regarding system design, maintenance and regeneration.

- Finally, pH variations within the CW-BES systems - which were statistically significantly different in this study when compared to CW-control - are suggested to be investigated further with equipment allowing for observations at the micro-scale near the cathode and anode.

**ACKNOWLEDGEMENTS**

This project has received funding from the European Union´s Horizon 2020 research and innovation programme under the Marie Skłodowska-Curie grant agreement No 676070. This communication reflects only the authors' view and the Research Executive Agency of the EU is not responsible for any use that may be made of the information it contains. Authors from GEMMA, UPC are grateful to the Government of Catalonia (Consolidated Research Group 2017 SGR 1029). M.J. García-Galán would like to thank the Spanish Ministry of Economy and Competitiveness for her Juan de la Cierva research grant (IJCI-2017-34601). Marianna Garfí is grateful to the Spanish Ministry of Economy and Competitiveness (Plan Estatal de Investigación Científica y Técnica y de Innovación 2013-2016, Subprograma Ramón y Cajal (RYC) 2016).

**SUPPLEMENTARY INFORMATION**

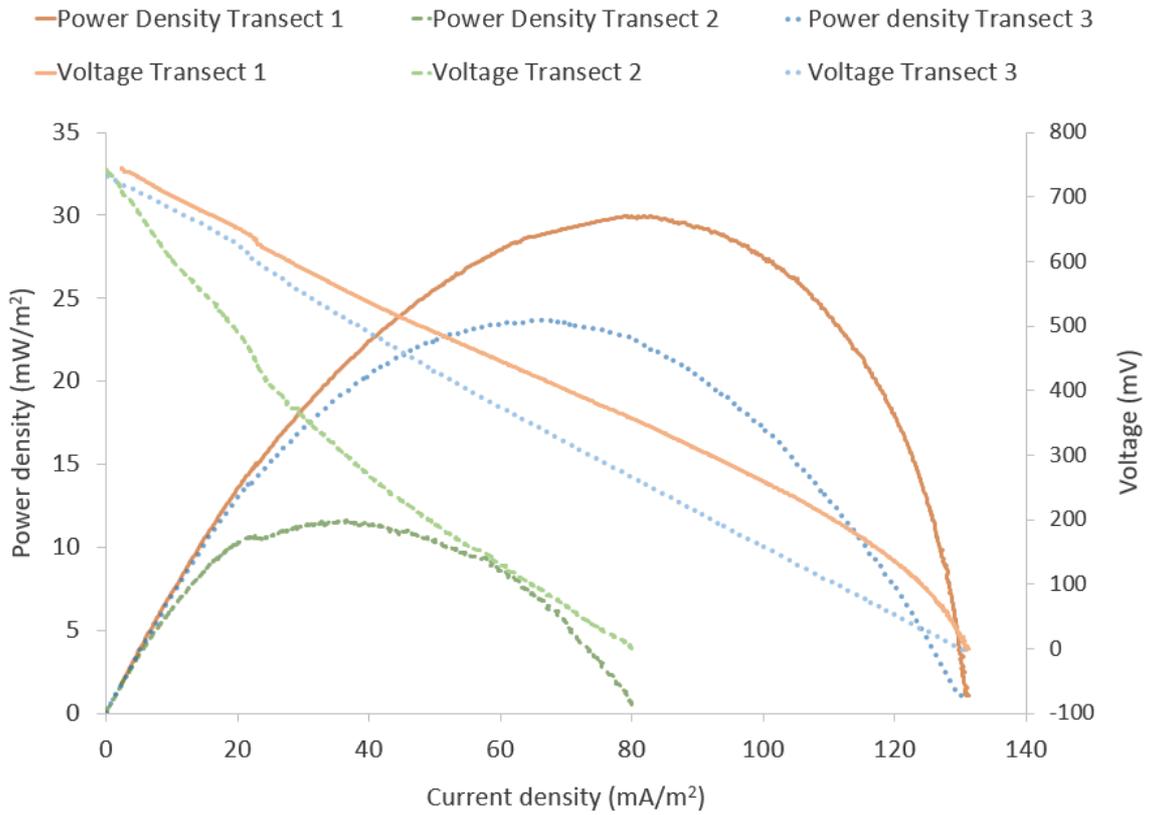

**Figure S1. Power density and polarization curves for each transect of one of the closed-circuit CW-MFC replicates measured during sampling week 4.**

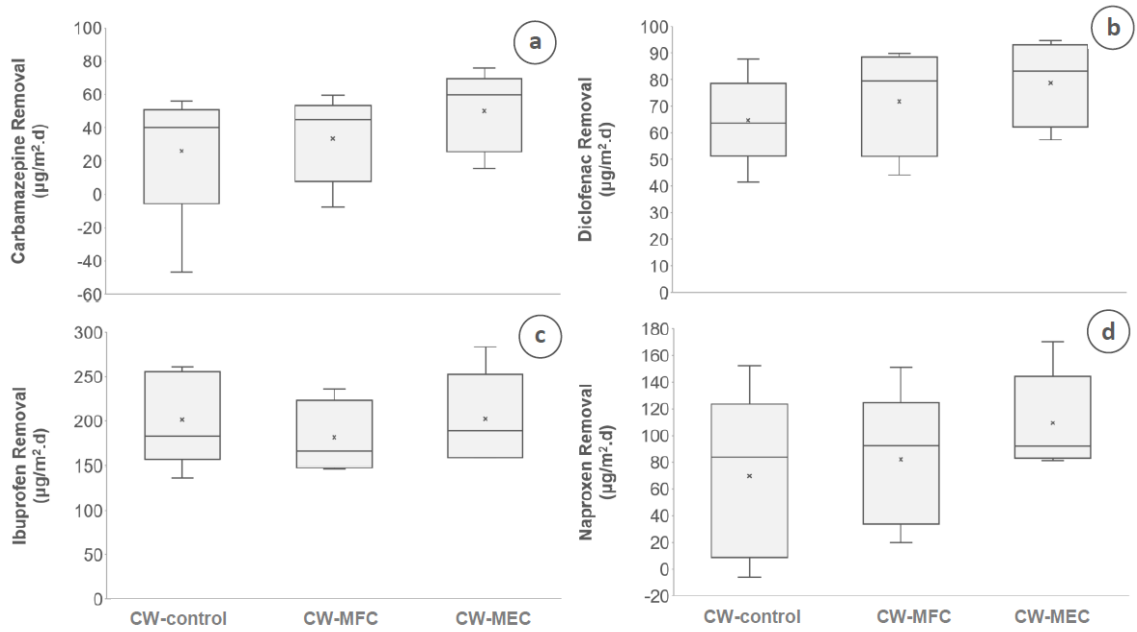

**Figure S2.** Specific removal from influent to effluent for all four OMPs (a; CBZ, b; DCF, c; IBU and d; NPX) comparing CW-control, CW-MFC and CW-MEC treatments (n=5). The box- and whisker plots show the minimum and maximum (lower and upper whiskers), first and third quartile (lower and upper end of box), median (horizontal line in box) and average (marked as an "x") values.

**Table S1.** Chemical structure and characteristics of the selected OMPs used in this study and their respective hydrophobicity and charge states estimated from the compound's Log D and $pK_a$, respectively (relative to the experimental pH of 7 – 7.5). Log $K_{ow}$ describes the octanol-water partition coefficient which is a compound's measure of the ratio of concentrations in octanol and water (Schwarzenbach et al., 2003). Log D is the partition coefficient for a compound at a specified pH

| Compound | Structure [a] | Classification | Log $K_{ow}$ | Log D (pH 6.6-7.6)[c] | Hydro-phobicity | $pK_a$[b] | Charge state |
|---|---|---|---|---|---|---|---|
| Carbamazepine | 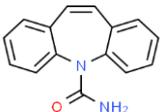 | Anticonvulsant | 2.45 [b] | 2.77 | hydrophilic | 13.90 | neutral |
| Diclofenac | 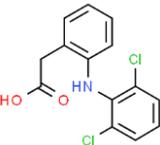 | Anti-inflammatory | 4.51 [d] | 1.70 to 1.04 | hydrophilic | 4.15 | negative |
| Ibuprofen | 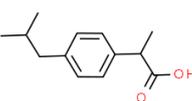 | Anti-inflammatory | 3.97 [b] | 2.10 to 1.16 | hydrophilic | 5.30 | negative |
| Naproxen | 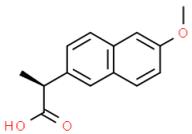 | Anti-inflammatory | 3.18 [b] | 0.61 to -0.18 | hydrophilic | 4.15 | negative |

[a] chemspider.com
[b] https://pubchem.ncbi.nlm.nih.gov
[c] chemicalize.com (data has been obtained from the empirical model)
[d] Avdeef et al. (1998)

**Table S2.** Results for pH for CW-control, CW-MFC and CW-MEC systems during the OMP spiking and sampling weeks at the influent, after first transect, after second transect and effluent as well as overall average.

|  |  | Influent | 1/3 | 2/3 | Effluent | Average |
|---|---|---|---|---|---|---|
| pH (-) | CW-control | 7.50±0.00 | 7.35±0.05 | 7.35±0.00** | 7.70±0.01 | 7.48±0.02 |
|  | CW-MFC | 7.45±0.05 | 7.09±0.02 | 7.05±0.07** | 7.66±0.07 | 7.32±0.05 |
|  | CW-MEC | 7.54±0.07 | 6.69±0.09** | 6.60±0.05** | 7.15±0.03** | 7.00±0.06 |

** ANOVA very significant difference (p < 0.01)

**Table S3.** Post-hoc Tukey HSD pairwise comparison results for pH in CW-control, CW-MFC and CW-MEC systems during the OMP spiking and sampling weeks after first transect, after second transect and effluent.

| | pH (-) Tukey HSD results | | | | | | | | |
|---|---|---|---|---|---|---|---|---|---|
| | 1/3 | | | 2/3 | | | Effluent | | |
| Pair | Q statistic | p-value | Inference | Q statistic | p-value | Inference | Q statistic | p-value | Inference |
| CW-control vs CW-MFC | 4.1187 | 0.06 | insignificant | 13.39 | 0.001 | ** p<0.01 | 0.586 | 0.9 | insignificant |
| CW-control vs CW-MEC | 10.6203 | 0.001 | ** p<0.01 | 33.51 | 0.001 | ** p<0.01 | 7.633 | 0.004 | ** p<0.01 |
| CW-MFC vs CW-MEC | 6.5016 | 0.009 | ** p<0.01 | 20.12 | 0.001 | ** p<0.01 | 7.047 | 0.006 | ** p<0.01 |

\* significant difference (p < 0.05)
\*\* very significant difference (p < 0.01)

**Table S4.** Post-hoc Scheffé pairwise comparison results for pH in CW-control, CW-MFC and CW-MEC systems during the OMP spiking and sampling weeks after first transect, after second transect and effluent.

| | pH (-) Scheffé results | | | | | | | | |
|---|---|---|---|---|---|---|---|---|---|
| | 1/3 | | | 2/3 | | | Effluent | | |
| Pair | TT-stats | p-value | Inference | TT-stats | p-value | Inference | TT-stats | p-value | Inference |
| CW-control vs CW-MFC | 2.9124 | 0.071 | In-significant | 9.468 | 2E-04 | ** p<0.01 | 0.414 | 0.919 | In-significant |
| CW-control vs CW-MEC | 7.5097 | 9E-04 | ** p<0.01 | 23.7 | 1.18E-06 | ** p<0.01 | 5.398 | 0.005 | ** p<0.01 |
| CW-MFC vs CW-MEC | 4.5973 | 0.011 | * p<0.05 | 14.23 | 2.39E-05 | ** p<0.01 | 4.983 | 0.007 | ** p<0.01 |

\* significant difference (p < 0.05)
\*\* very significant difference (p < 0.01)

Table S5. One-factor ANOVA (with replication) results for the comparison of conventional wastewater parameters between the electric connections during the sampling period, for the total system from inlet to outlet and each of the three transects separately (statistically significant different if p-value < 0.05).

| One-factor ANOVA | | p-value Comparing Electric Connections | | | |
|---|---|---|---|---|---|
| | | Inlet-Outlet | Transect 1 | Transect 2 | Transect 3 |
| COD | $F_{(2, 8)}$ | 0.37 | 0.84 | 0.42 | 0.97 |
| $NH_4$-N | $F_{(2, 7)}$ | 0.20 | 0.21 | 0.93 | 0.99 |
| $SO_4^{2-}$ | $F_{(2, 6)}$ | 0.97 | 0.98 | 0.16 | 0.36 |
| $PO_4$-P | $F_{(2, 6)}$ | 0.96 | 0.76 | 0.57 | 0.20 |

Table S6. One-factor ANOVA (with replication) results for the comparison of the four tested OMPs between the electric connections during the sampling period, for the total system from inlet to outlet and each of the three transects separately (statistically significant different if p-value < 0.05).

| One-factor ANOVA | | p-value Comparing Electric Connections |
|---|---|---|
| | | Inlet-Outlet |
| Carbamazepine (CBZ) | $F_{(2, 5)}$ | 0.48 |
| Diclofenac (DCF) | $F_{(2, 5)}$ | 0.48 |
| Ibuprofen (IBU) | $F_{(2, 5)}$ | 0.75 |
| Naproxen (NPX) | $F_{(2, 5)}$ | 0.47 |